\documentstyle[epsf]{mn}
\ifnfsstwo

\fi
 
\ifnfssone
  \newmathalphabet{\mathit}
    \addtoversion{normal}{\mathit}{cmr}{m}{it}
    \addtoversion{bold}{\mathit}{cmr}{bx}{it}

\fi

\ifoldfss

\fi

\loadboldmathitalic
\loadboldgreek
\title[Clustering of the Lyman-$\alpha$ clouds]
{The clustering properties of the Lyman-$\alpha$ clouds}

\author[S. Cristiani et al.]
       {S. Cristiani$^{1,3}$,
 S. D'Odorico$^1$, V. D'Odorico$^{2,3}$, 
 A. Fontana$^4$, E. Giallongo$^4$, \and S. Savaglio$^1$ \\
  $^1$European Southern Observatory, K. Schwarzschild Strasse 2,
	D-85748 Garching, Germany \\
  $^2$International School for Advanced Studies, SISSA, via Beirut 2-4,
	I-34013 Trieste, Italy \\
  $^3$Dipartimento di Astronomia, Universit\`a di Padova,
    vicolo dell'Osservatorio 5, I-35122 Padova, Italy\\
  $^4$Osservatorio Astronomico di Roma, via dell'Osservatorio,
I-00040, Monteporzio, Italy \\
  }

\date{Accepted \dots 
      Received \dots ;
      in original form 1995 December 28}
\pubyear{1995}
\begin{document}
\maketitle
\begin{abstract}
We analyze the clustering properties of a high-resolution ($\sim
10$ km/s) sample of about 1600 Lyman-$\alpha$ lines in the
spectra of 15 quasars, obtained in the framework of an ESO
key-programme with the addition of literature data. The two-point
correlation function in the velocity space shows a significant signal
on small velocity scales ($\Delta v \la 300$ km/s) with amplitude and
significance increasing with increasing column density. The
correlation scale at $z \sim 3$ turns out to be $200 - 300 h_{50}^{-1}$ kpc.
A trend of increasing correlation with decreasing redshift is
apparent. The existence of over- and under-densities on scales of a
few tens of Megaparsec is confirmed with a high confidence level and a
number of possible structures are identified. The present observations
are found to be consistent with models  of gravitationally induced
correlations. A continuity scenario between  Lyman-$\alpha$ and metal
systems emerges,
with a suggested physical 
association between the Lyman-$\alpha$ clouds with $\log N_{HI}
\ga 14$  and the halos of protogalactic systems.

\end{abstract}


\section{Introduction}

The Lyman-$\alpha$ ``forest'' of absorption lines seen in the spectra of
high-redshift quasars is generally ascribed to pre-galactic or proto-galactic
material. The study of the spatial distribution of these clouds offers a
powerful tool to gain insight in the formation and evolution of structures up
to redshifts 4-5, much larger than those reachable with observations of
galaxies, limited at $ z \la 1$. 
The only other class of objects that at present can provide this type of
information are the QSOs themselves. Lyman-$\alpha$ absorbers have,
however, two definite advantages with respect to QSOs: they are much more 
numerous and, being less exotic objects, are expected to trace more ``normal''
matter.

The search for clustering of the Lyman-$\alpha$ lines has provided
along the years results that have been defined disappointing or even 
controversial. 
Systematic studies of the distribution of redshifts in the QSO Lyman-$\alpha$
forest began in the early 1980's with the work by Sargent et al.
\shortcite{sarg80},
which concluded that no structures could be identified. 
Almost all the subsequent results have failed to detect any significant 
correlation on velocity scales
$300 < \Delta v < 30000$ km s$^{-1}$ \cite{sarg82,becht87,webb91}. 
On smaller scales ($\Delta v =50-300$ km s$^{-1}$) 
there have been indications of weak clustering 
\cite{webb87,rauch92,cherno95,cris95}, together with 
relevant non-detections \cite{pet90,eric93}.

On the contrary, metal-line systems selected by means of the CIV doublet
\cite{sarg88} have been early recognized to show
strong clustering up to 600 km s$^{-1}$, suggesting a different spatial
distribution.
In fact, the absence of power in the two-point correlation function
has been claimed as a striking characteristics of the Lyman-$\alpha$ forest
and has been used as a basic argument to develop a scenario of the
Lyman-$\alpha$ absorbers as a totally distinct population with respect to
metal systems and therefore galaxies.

\section{The database}

The present analysis is based on data obtained in the framework of an ESO
key-programme devoted 
to the study of QSO absorption systems at high redshifts.
Up to now spectra of six QSOs, 
at resolutions between 9 and 14 km s$^{-1}$, have been reduced.
The list of the objects, with emission redshifts ranging from 3.27 to
4.12, is given in Table 1 together with the main characteristics
of the sample of Lyman-$\alpha$ absorption lines. 
This constitutes a unique database, especially at high redshift, where
the density of lines provides particular sensitivity to any clustering signal.

\begin{table}
 \caption{QSO Spectra from the ESO KP}
 \label{tab1}
 \begin{tabular}{cccccc}
\hline
\hline
QSO  & $z_{em}$ & FWHM          & Mag & No of & Sample \\
Name &          & (km s$^{-1}$) &     & lines & Limit \\
     &          &               &     &       & $\log(N_{HI}$) \\
\hline
\ \\
$2126-15$   & 3.27 & 11 & V=17.3 & 106 & 13.3 \\
$2355+01$   & 3.39 & ~9 & V=17.5 & ~76 & 13.3 \\
$0055-26$   & 3.67 & 14 & V=17.5 & 187 & 13.3 \\
$1208+10$   & 3.82 & ~9 & V=17.5 & ~66 & 13.3 \\
$1108-07$   & 3.95 & ~9 & R=18.1 & ~38 & 13.3 \\
$0000-26$   & 4.12 & 12 & V=17.5 & 181 & 13.3 \\
\end{tabular}

\bigskip\bigskip

 \begin{tabular}{cccccc}
\multicolumn{6}{l}{Other QSO Spectra from the literature}
\ \\
\hline
\hline
QSO  & $z_{em}$ & FWHM          & Mag & No of & Sample \\
Name &          & (km s$^{-1}$) & (V) & lines & Limit \\
     &          &               &     &       & $\log(N_{HI}$) \\
\hline
\ \\
$1331+17^a$ & 2.10 & 18 & 16.9 & ~55 & 13.0 \\
$1101-26^b$ & 2.15 & ~9 & 16.0 & ~37 & 13.3 \\
$2206-19^c$ & 2.56 & ~6 & 17.3 & ~47 & 13.3 \\
$1700+64^d$ & 2.72 & 15 & 16.1 & ~73 & 13.3 \\ 
$1946+76^e$ & 3.02 & 10 & 15.8 & ~72 & 13.3 \\
$0636+68^f$ & 3.17 & ~8 & 16.6 & 172 & 13.0 \\
$0302-00^f$ & 3.29 & ~8 & 17.6 & 151 & 13.0 \\
$0956+12^f$ & 3.30 & ~8 & 17.6 & 151 & 13.0 \\
$0014+81^f$ & 3.41 & ~8 & 16.5 & 164 & 13.0 \\
\ \\
\multicolumn{4}{l}{References:}\\
\multicolumn{4}{l}{\ $^a$ Kulkarni et al. 1995}\\
\multicolumn{4}{l}{\ $^b$ Carswell et al. 1991}\\
\multicolumn{4}{l}{\ $^c$ Rauch et al. 1993}\\
\multicolumn{4}{l}{\ $^d$ Rodr\'iguez-Pascual et al. 1995}\\
\multicolumn{4}{l}{\ $^e$ Fan and Tytler 1994}\\
\multicolumn{4}{l}{\ $^f$ Hu et al. 1995}\\
\end{tabular}
\end{table}

Spectra of the QSOs $2126-158$, $0055-269$, $0000-2619$ have been already
published \cite{gial93,cris95,sava96}.
For $2126-158$ new spectra have been added, significantly improving
the S/N with respect to the published data.
A complete description of the individual spectra will be given 
elsewhere: 
here we discuss the clustering properties of the Lyman-$\alpha$
sample.

All the spectra have been analyzed in a uniform way and all the lines have
been fitted using our FITLYMAN program which is now available in the ESO-MIDAS
package \cite{adri95}.
It performs a $\chi^2$ minimization to derive the redshift $z$,
the Doppler parameter $b$ and the column density $N$ for
isolated lines and individual components of the blends.
Whenever possible, the Lyman-$\beta$ lines have been used to constrain
the number of components in the strong saturated Lyman-$\alpha$
blends. 

The S/N has been computed from the noise spectrum, and is typically
greater than
10 per pixel element (corresponding to 15 per resolution element). 
In the regions near the QSO 
Lyman-$\alpha$ emissions
and for the brighter quasars
the S/N raises to $\sim 20-60$ per resolution element.
Extensive simulations by various authors
\cite{rauch93,adri95}
show that no strong bias in the derived statistical
distributions is expected when the S/N is $>$10.

To identify heavy element systems we have compared the lists of the
observed lines with a catalog, derived from Morton (1991), containing the 
most frequently seen lines in QSO absorption spectra.
As customary, we   
searched for significant excesses of identifications at all possible
redshifts \cite{Bahc68}.

We have complemented our data with other spectra available in the 
literature with similar resolution and redshift range,
obtaining a final sample of 15 QSOs (in the following {\it extended
sample}, see Table 1).

All the identified heavy element systems present in the Lyman-$\alpha$ forest 
have been removed from the extended sample, which consists of almost 
1600 Lyman-$\alpha$ lines with observed
column density $\log N_{HI}\geq 13$.

Lines affected by the proximity effect, i.e. within a proper distance
of $10$ Mpc from the emission redshift of each QSO,
have been excluded from the clustering analysis.

We adopt throughout the value H$_0 = 50$ km s$^{-1}$
Mpc$^{-1}$ for the Hubble parameter and $q_0 = 0.5$. 

\section{The statistical analysis of the clustering}

\subsection{Small scales}

As a first statistical tool to study the clustering properties of our
sample of  Lyman-$\alpha$ lines we have adopted
the two-point correlation function (TPCF),
defined as the excess, due to clustering, of the probability $dP$
of finding a Lyman$-\alpha$ cloud in a volume $dV$ at a distance $r$ from 
another cloud:
\begin{equation}
dP=\Phi _{Ly\alpha}(z) dV [1+\xi(r)]
\end{equation}
where $\Phi (z)$ is the average space density of the clouds as a function of
$z$. 
The TPCF is known to be a satisfactory
estimator when used to investigate weak clustering on 
scales considerably smaller than
the total interval covered by the data, which is the first domain we
want to explore. The binning, intrinsic
to this method, causes a loss of information, but the 
ease in visualizing its results and in including observational effects
in the computing codes have made of the TPCF
one of the favorite statistical estimators in cosmology.

In practice the observations provide the redshifts of the Lyman-$\alpha$
lines that, due to peculiar motions, are not immediately transformed 
in comoving distances. Therefore it is normally preferred to compute
the TPCF in the velocity space, estimated as
\cite{Peeb80}
\begin{equation}
\xi(\Delta v) = {N_{obs}(\Delta v) \over N_{exp}(\Delta v)} -1
\end{equation}
where $N_{obs}$ is the observed number of line pairs with velocity separations
between $\Delta v$ and $\Delta v + \epsilon_v$ and $N_{exp}$ 
is the number of pairs
expected in the same interval from a random distribution in redshift.

At the small velocity separations we are dealing with, the
variation of the distance scale with cosmic time can be neglected
and the velocity difference can be simply deduced from the redshift difference
\cite{sarg80}
\begin{equation}
\Delta v = {c ~ (z_2-z_1) \over 1 + (z_1+z_2)/2}
\end{equation}
where $\Delta v$ is the velocity of one cloud as measured by an observer in
the rest-frame of the other.

\begin{figure}
\epsfxsize=85truemm
\epsffile{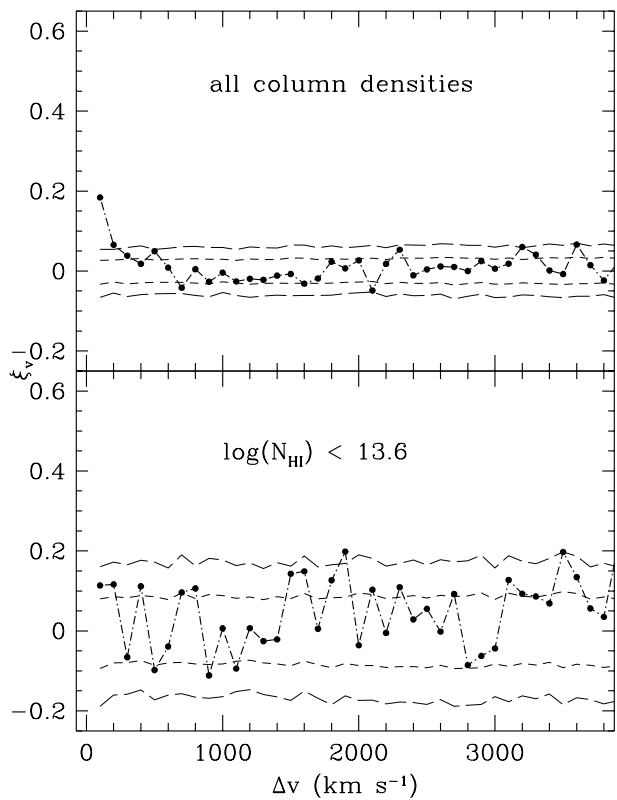}
 \caption{Two-point correlation function in the velocity space: $a)$
(upper panel)
for the complete sample of Lyman-$\alpha$ lines, $b)$ (lower panel)
for lines with
column densities $< 10^{13.6}$ cm$^{-2}$. The short-dashed and
long-dashed lines represent the $1 \sigma$ and $2 \sigma$ confidence
limits for a Poissonian process}
 \label{fig1}
\end{figure}

In our line sample $N_{exp}$ is obtained averaging 1000 numerical simulations
of the observed number of redshifts, trying to account for all the relevant
cosmological and observational effects.
In particular the set of redshifts is randomly generated in the same
redshift interval as the data according
to the cosmological distribution $\propto (1+z)^{\gamma}$, where the
best value of $\gamma = 2.65$ has been derived from a
maximum likelihood analysis of the same database \cite{gial96}.
The results are not sensitive to the value of $\gamma$ adopted and
even a flat distribution (i.e. $\gamma =0$) gives values of $\xi$ that
differ typically by less than 0.02.
Incomplete wavelength coverage due to gaps in the spectrum or line blanketing
of weak lines due to strong complexes is also accounted for.
Lines with too small velocity splittings,
compared with the finite resolution or the
intrinsic blending due to the typical line widths -
the so called ``line-blanketing'' effect \cite{gial96} -,
are excluded in the estimate of $N_{exp}$.

\begin{figure}
\epsfxsize=85truemm
\epsffile{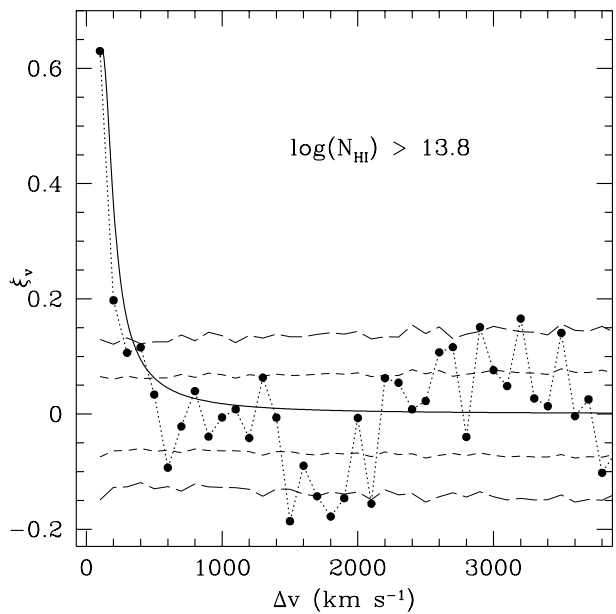}
 \caption{Two-point correlation function in the velocity space 
 for lines with
 column densities $> 10^{13.8}$ cm$^{-2}$. Confidence limits as in
Fig. 1.
The continuous line shows the model described in Section 4, eq. 4,
with $\gamma = 1.77$, $\sigma = 50$ km/s, $r_{cl}= 180 h_{50}^{-1} $
kpc and $r_0 =
250 h_{50}^{-1}$ kpc at $ z = 3$.
}
 \label{fig2}

\end{figure}

The resulting correlation function for the full {\it extended
sample} of Lyman$-\alpha$ lines
is shown in the upper panel of Fig. 1. 
A weak but significant signal is present with $\xi\simeq 0.2$  in the
100 km s$^{-1}$ bin: 739 pairs are observed while only 624 are
expected for a random distribution, a $4.6 \sigma$ deviation from
poissonianity. 

As in Cristiani et al. (1995), we have explored the variations of the
clustering as a function of the column density.
In Fig. 1b the correlation function for lines with $\log N_{HI}\le 13.6$
is shown. All the evidence for clustering has disappeared.
On the contrary, for lines with $\log N_{HI}\ga 13.8$ (Fig. \ref{fig2}),
the correlation function at $\Delta v = 100$ km s$^{-1}$ 
shows a remarkable increase in amplitude ($\xi\simeq 0.6$) and
significance: 234 pairs are observed while only 145 are
expected for a random distribution, a more than $7 \sigma$ deviation from
poissonianity. 
No relevant feature other than the peak at small velocity separations
is observed. In particular the evidence for anti-clustering on scales
$\sim 600 - 1000$ km s$^{-1}$, previously
suggested by Meiksin and Bouchet \shortcite{meiks95} on the basis of
data of $0055-259$ and $0014+813$, is not confirmed.
It should be noted that the assessment of the significance of a
deviation from poissonianity is conceptually different when
a velocity bin {\it given a priori} 
(as in the analysis of small-scale clustering)
is analyzed, 
with respect to the case in which
a deviation in {\it any} of the bins is searched for.
While in the former situation the confidence levels shown in Figg.
1,2,4 (obtained  by comparing the number of pairs observed in each bin
with the distributions expected from simulations) are perfectly valid,
in the latter case the significance
of a positive or negative deviation in {\it any} of the bins is
considerably less than what a naive comparison 
would at first glance suggest.

\begin{figure}
\epsfxsize=85truemm
\epsffile{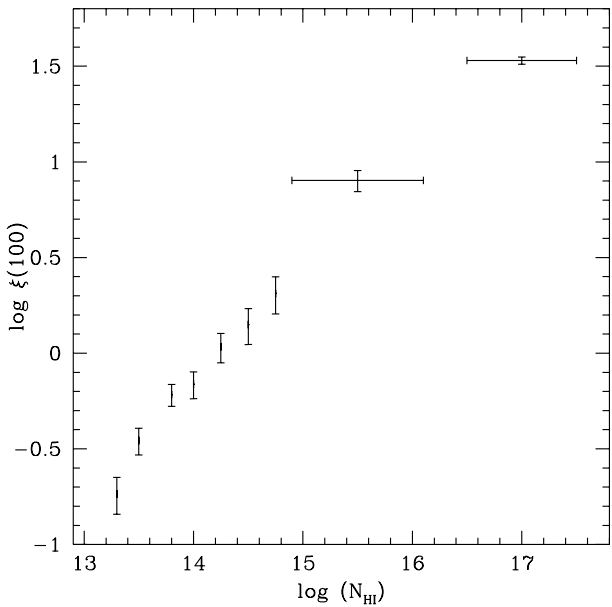}
 \caption{Evolution of the amplitude of the two-point correlation
function as a function of the column density threshold for the sample
of the Lyman-$\alpha$ lines. The two points in the upper-right side of the 
picture show the correlation of the CIV metal systems (see text)}
 \label{figclustew}
\end{figure}

Fig.~\ref{figclustew} shows the variation of the amplitude of the
two-point correlation as a function of the column density threshold
for the sample of the Lyman-$\alpha$ lines. 
On the same plot are reported for comparison the TPCF's
derived for high and low column density CIV metal systems 
from the works by Petitjean and Bergeron \shortcite{CIVpetit}
and by Songaila and Cowie \shortcite{son:cow:96}, respectively. 
A reference value of $\log (N_{HI}) \sim 17 \pm 0.5  $  has been
assigned to high column density CIV metal systems on the basis of observations
of 0420-388 \cite{ref0420,CIVpetit} and of 2126-158
\cite{gial93,CIVpetit}. For the low column density CIV systems a value
of $\log (N_{HI}) \sim 15.5 \pm 0.6 $ is derived from Songaila and
Cowie \shortcite{son:cow:96}.

An extrapolation of the increasing amplitude trend
observed for the TPCF of the Lyman-$\alpha$ lines
would easily intercept the corresponding estimates derived from the CIV metal
systems. We stress also the similarity between the shapes
of the TPCF's of Lyman-$\alpha$ and
CIV systems, when observed at comparable resolution
(cfr. Fig. 4b of Petitjean and Bergeron 1994, Womble et al. 1995,
Fig.6 of Songaila and Cowie 1996) 
and the fact that a trend of
increasing correlation with increasing column density is observed 
also in CIV systems, for which the strong systems appear usually in 
clumps of many components, while the weak doublets (generally optically
thin at the Lyman limit) are typically made of an isolated, single
component \cite{berg95}.

In a recent paper, Fernandez-Soto et al. (1996) have investigated the
clustering properties of the Lyman-$\alpha$ clouds on the basis of the
corresponding CIV absorptions, suggesting that CIV may resolve better
the small-scale velocity structure that cannot be fully traced by 
Lyman-$\alpha$ lines. As a consequence, the estimates of the TPCF of 
the Lyman-$\alpha$ absorbers, although the effects of the non-negligible
width of the lines (the ``line-blanketing'')
are considered in our simulations,
should be regarded as a lower limit to the real clustering amplitude.
However, it is not straightforward to translate the properties of the
CIV absorbers in the corresponding ones of the 
Lyman-$\alpha$ absorbers, since observations (e.g. 
Prochaska and Wolfe 1996) show that the velocity structures of
high and low-ionization species are often different.
Since the underestimation of the Lyman-$\alpha$ TPCF would be more severe at
larger column densities, the trend of an increasing correlation with
increasing column density appears to be real, from the lowest column
densities up to those corresponding to the strongest metal systems.

\begin{figure}
\epsfxsize=85truemm
\epsffile{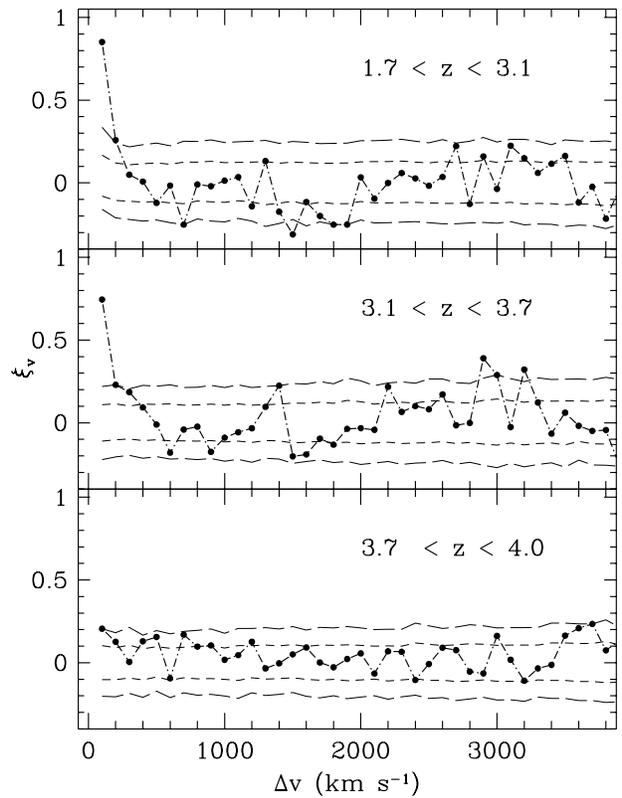}
 \caption{Evolution of the two-point correlation
function with redshift for the sample
of the Lyman-$\alpha$ lines with column densities above
$\log(N_{HI})=13.8$. The three redshift intervals  $1.7<z<3.1$,
$3.1<z<3.7$, $3.7<z<4.0$ (upper, middle and lower panel, respectively)
have been chosen in a way that the confidence limits, estimated as in
Fig. 2-3, turn out to be  roughly the same in all three ranges}
 \label{figclustewz}
\end{figure}

In Fig. \ref{figclustewz} we study the evolution of the TPCF 
with the redshift for the sub-sample
of Lyman-$\alpha$ lines with column densities $\log(N_{HI}) > 13.8$. 
The amplitude of the correlation at $100$ km s$^{-1}$ decreases
with increasing redshift from $0.85\pm0.14$ at $1.7<z<3.1$, to
$0.74\pm0.14$ at $3.1<z<3.7$ and $0.21\pm0.14$ at $3.7<z<4.0$.
Unfortunately, HST data are still at too low-resolution (e.g. Bahcall et
al. 1995) or are too scanty \cite{3C273} to allow a meaningful comparison
with the present data. Nonetheless, the existence at
low-redshifts of clumps of Lyman-$\alpha$ absorption lines clustered 
around metal-line systems, reported by Bahcall et al. (1995) as a
resemblance with the properties of galaxies and of metal absorption
systems, is suggestive that the trend illustrated in
Fig.~\ref{figclustewz} persists at lower redshifts, as confirmed by the 
measure of the Lyman-$\alpha$ TPCF, $\xi = 1.8^{+1.6}_{-1.2}$, on
scales of $200-500$ km/s at $0 < z < 1.3$  \cite{ulmer}.

\subsection{Large scales}

The TPCF is known to be inefficient for
detecting structures on scales comparable with respect to the
dimension of the sample (e.g. Pando and Fang 1996).
For this reason to investigate the clustering properties of the
Lyman-$\alpha$ absorbers on scales of several comoving megaparsecs, 
searching for over- and under-densities of
lines significantly deviating from a Poissonian distribution, 
we have resorted to different statistical techniques.
The aim was not only to assess the existence of non-random
fluctuations on large scales, but also to identify the regions of the
spectra giving origin to such deviations.

\subsubsection{Voids}

A typical way of looking for non-random fluctuations of the line
density on large scales is the search for voids.
Voids in the Lyman-$\alpha$ forest provide a test for models
of the large-scale structure in the Universe and of the homogeneity
of the UV ionizing flux. Previous 
searches for megaparsec-sized voids have produced a few claims
\cite{Crotts89,DB91}, but uncertainties in the line statistics
strongly influence the probability estimate \cite{Ostriker88}. 
High resolution data, less affected by blending effects, are ideal
also for the study of voids, but great care has to be taken
in the statistical
approach to avoid the pitfalls of ``a posteriori statistics''.
Detections of voids in individual cases are interesting but 
a more general approach, assessing more quantitatively
how common is the phenomenon, is to be preferred.
Differences in the redshifts and S/N (and
consequently in the line density) make difficult a global and uniform usage
of our extended sample in this sense.
Therefore we have designed the following procedure:
for each of the objects listed in Table \ref{tab1} and for lines with
$\log (N_{HI}) \geq 13.3$ we have
estimated through Montecarlo simulations the void size $\Delta r$ (in comoving
coordinates) for which the
probability to find at least one void $ \ge \Delta r$
is $0.05$. The objects for which the density of lines and the absence
of spectral gaps allow a meaningful analysis 
are listed in Table \ref{tabvoid}.
Then we have searched each spectrum for voids of dimensions
equal or larger than $\Delta r$. In 3 out of 9 cases at least
one void was detected (two voids for $0055-26$). The binomial
probability corresponding to such occurrences is $8 \cdot 10^{-3}$.

\begin{table}
 \caption{QSO Spectra used in the search for voids}
 \label{tabvoid}
 \begin{tabular}{cccc}
\hline
\hline
QSO  & $z_{em}$ & minimum   & Number of \\
Name &          & void size & detected  \\
     &          & $\Delta r$ (Mpc)     & voids     \\
\hline
\ \\
$0636+68$   & 3.17 & 15.9 & 1 \\
$2126-15$   & 3.27 & 24.7 & 0 \\
$0302-00$   & 3.29 & 17.5 & 0 \\
$0956+12$   & 3.30 & 16.9 & 0 \\
$2355+01$   & 3.39 & 13.1 & 0 \\
$0014+81$   & 3.41 & 15.8 & 0 \\ 
$0055-26$   & 3.67 & 14.9 & 2 \\ 
$1208+10$   & 3.82 & ~8.5 & 0 \\
$0000-26$   & 4.12 & ~6.9 & 1 \\
\hline
\hline
\end{tabular}
\end{table}

The four voids reported in Table \ref{tabvoid} have been found in the
spectra of $0636+68$ at $4339.5 - 4361.6$ \AA\ ($16.1$ Mpc size), 
$0055-26$
at $4979.05 - 5016.6$ and $5188.4 - 5223.8$
\AA\ ($22.3$ and $19.8$ Mpc, respectively)
and $0000-26$ at $5627 - 5644$ \AA\ ($8.5$ Mpc).
The well known void at $5050 - 5100$ \AA\ in the spectrum of $0302-00$
\cite{DB91} has not
been found because the spectrum of Hu et al. (1995) does not cover this
region of the Lyman forest. 
From the published spectra and from Fig. 5 it is apparent that the regions
corresponding to the voids are not completely devoid of lines: weak
absorptions are observed within the voids. This agrees with
low-redshift observations \cite{shull96} that have shown that in the
local Universe voids are not entirely devoid of matter.

Even if underdense regions are statistically significant in our sample,
the filling factor is rather low: the voids reported in Table
\ref{tabvoid} cover only about $2$\% of the available line-of-sight
path-length, confirming what already noticed by Carswell and Rees
(1987).

\begin{figure*}
\epsfxsize=180truemm
\epsffile{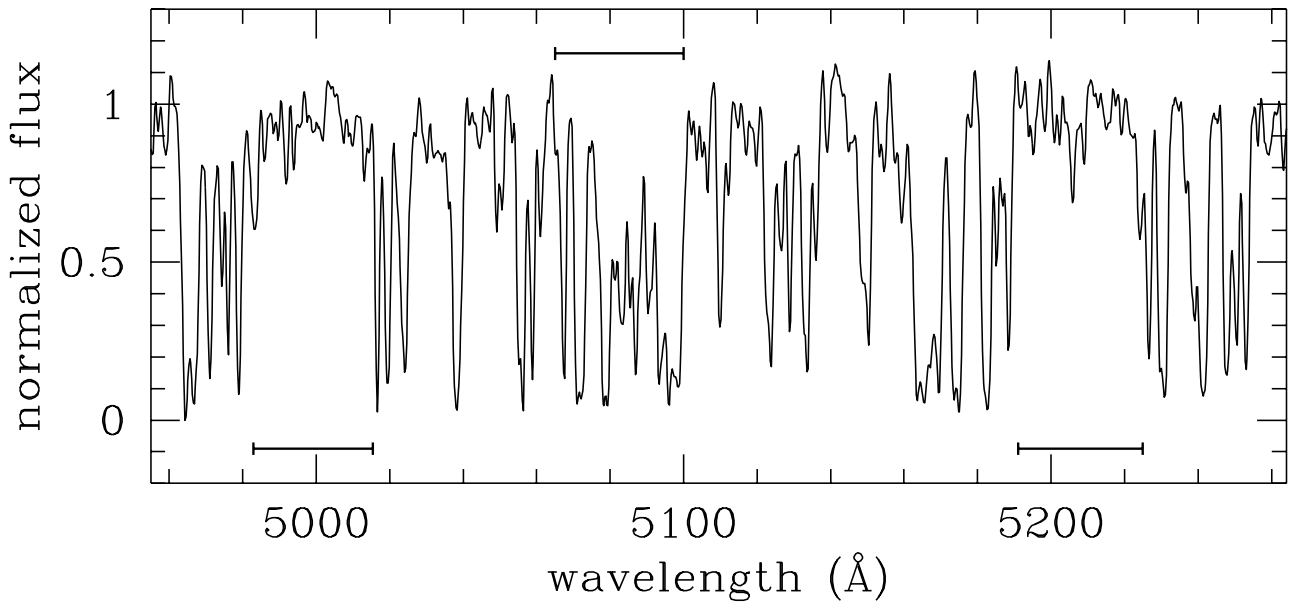}
 \caption{ Over and under-densities of lines at a scale of 20 Mpc
in the spectrum of
Q$0055-26$. The positions of two voids, at 5000 and 5200 \AA\
respectively (see Tab. 2), are shown together with and
over-density at 5080 \AA. The relatively low resolution of the plot
does not allow to fully visualize the significance of the over-density:
15 lines with $\log (N_{HI}) \geq 13.3$ are observed while only 7 are
expected.}
 \label{fig0055}
\end{figure*}

\subsubsection{Over- and under-densities of lines}

Voids are just an extreme case of spectral regions showing an
under-density of lines. Various theoretical reasons prompt to tackle 
the issue
of over- and under-densities of lines
from a more general point of view: the typical signature of a
``proximity effect'' due to a foreground quasar is a lack of weak
lines in the Lyman forest, rather than lack of lines in general
\cite{DBVict}; the relative filling factor of under-densities and
over-densities 
may provide a constraint on the theories of structure formation.

We have analysed the spectra of the quasars reported in Tab.
\ref{tab1} with a counts-in-cells technique
searching for over- and under-densities of lines with $\log (N_{HI})
\geq 13.3 $ on scales from 10 to 80 Mpc and comparing the observed counts
with Montecarlo simulations in order to assess the significance of the
deviations. 
On smaller scales the shot noise is too
large, on larger scales the ``integral constraint'', forcing the
simulated number of lines to be equal to the observed one, prevents
from the possibility of detecting any deviation.
The threshold to define significant a deviation (for example in excess)
in a given spectral 
interval has been set in a way that at a
given scale for a given quasar there is a $0.05$ probability
of observing 
at least one deviation of this type on the whole spectrum for a
locally Poissonian distribution.
In Table \ref{tabdensi} we report the results of the analysis.
An asterisk denotes that a meaningful search at that particular scale has not
been possible for the given object.

\begin{table}
 \caption{Over-densities of lines found in QSO spectra}
 \label{tabdensi}
 \begin{tabular}{cccccc}
\hline
\hline
Scale & 10 Mpc & 20 Mpc & 30 Mpc & 40 Mpc & 80 Mpc  \\
\hline
QSO  & \multicolumn{5}{c}{over-density center} \\
Name & \multicolumn{5}{c}{(\AA)} \\
\hline
\ \\
$1331+17$   & 3494 & 3493 & 3490 & 3490 & -    \\
            & 3605 & 3609 \\
$1101-26$   & 3742 & -    & 3743 & -    & -    \\
$2206-19$   & -    & -    & -    & -    & -    \\
$1700+64$   & -    & -    & -    & -    & -    \\
$1946+76$   & 4427 & 4435 & -    & -    & -    \\
$0636+68$   & -    & -    & -    & -    & -    \\
$2126-15$   & -    & -    & -    & -    & -    \\
$0302-00$   & -    & -    & 4696 & -    & -    \\
$0956+12$   & -    & -    & -    & -    & -    \\
$2355+01$   & -    & 5217 & -    & -    & -    \\
$0014+81$   & -    & -    & -    & -    & -    \\
$0055-26$   & 5085 & 5080 & -    & -    & -    \\
$1208+10$   & 5621 & -    & 5645 & -    & $*$  \\
$1108-07$   & -    & -    & -    & 5839 & $*$  \\
$0000-26$   & -    & -    & -    & -    & -    \\
\hline
\hline
\end{tabular}
\end{table}

\begin{table}
 \caption{Under-densities of lines found in QSO spectra}
 \label{tabunder}
 \begin{tabular}{cccccc}
\hline
\hline
Scale & 10 Mpc & 20 Mpc & 30 Mpc & 40 Mpc & 80 Mpc  \\
\hline
QSO  & \multicolumn{5}{c}{under-density center} \\
Name & \multicolumn{5}{c}{(\AA)} \\
\hline
\ \\
$1331+17$   & *    & *    & 3707 & -    &  -   \\
$1101-26$   & *    & *    & 3679 & -    & 3670 \\
$2206-19$   & *    & 4238 & -    & -    & -    \\
$1700+64$   & *    & -    & -    & -    & -    \\
$1946+76$   & *    & -    & -    & 4799 & -    \\
$0636+68$   & *    & -    & -    & -    & -    \\
$2126-15$   & *    & -    & -    & -    & -    \\
$0302-00$   & *    & -    & -    & -    & -    \\
$0956+12$   & *    & -    & -    & -    & -    \\
$2355+01$   & *    & -    & -    & -    & -    \\
$0014+81$   & *    & -    & -    & -    & -    \\
$0055-26$   & 5263 & 5001 & 5007 & 4988 & -    \\
            &      & 5208 & 5406 &      &      \\
$1208+10$   & -    & -    & -    & -    & -    \\
$1108-07$   & -    & -    & -    & -    & $*$  \\
$0000-26$   & 5637 & -    & -    & -    & -    \\
\hline
\hline
\end{tabular}
\end{table}

5 QSOs out of 15 show at least one over-density in their spectrum at
10 Mpc scales,
corresponding to a binomial probability of $6 \cdot 10^{-4}$
of being drawn from a poissonian distribution of lines.
4 QSOs 
show at least one over-density of lines at 20 and 30 Mpc scales,
corresponding to a binomial probability of $5 \cdot 10^{-3}$.
At larger scales the number of significant over-densities decreases.
None is observed at 80 Mpc.
Over-densities detected at one scale tend to persist at larger scales,
analogously to what observed in the wavelet identification of structures
from galaxy counts \cite{slez90}.

Under-densities appear to be roughly as common as over-densities, 
once the lack of sensitivity 
at lower 
redshifts and smaller scales, due to the low density of lines, is
taken into account. 
In practice in Table \ref{tabunder} the voids of Table \ref{tabvoid} are
recovered, except for the void of $0636+68$, 
the least significant case.

\section{Discussion}

As shown by Fernandez-Soto et al. (1996), a TPCF approach applied
- as in the present work - directly to the Lyman-$\alpha$ lines
tends to underestimate their clustering properties.
However, the very fact that a signal is found, in spite of all the
limitations described by Fernandez-Soto et al. (1996), is a
significant result and, even if regarded as lower limits, the measured
clustering amplitudes allow to draw interesting inferences.

In this way,
the observed clustering properties are qualitatively consistent with a
scenario of gravitationally induced correlations.
The dependence of the clustering amplitude on the column density,
reported in subsection 3.1, is easily
explained by models involving biasing for the formation of structures
in the universe. Objects associated with the stronger potential wells 
are expected to be more clustered and the HI column density is
naturally related to the depth of the well (and to the associated
mass).

The trend of increasing correlation with decreasing redshift
(Fig. \ref{figclustewz}) is also a strong prediction of any model of
structure formation based on gravitational instability.
On the contrary, theories of explosive structure formation
\cite{vish85,wein89} expect a velocity correlation
either unchanging or diminishing with time.

The existence of a roughly equal number of over and under-densities
on scales $10 - 40$ Mpc is easily understandable in terms of the linear
theory of the evolution of the perturbations, that is a plausible
approximation at such relatively high redshifts: gravity has
not yet had time to give a significant skewness to the (under)over-density
distribution. Besides, almost any hierarchical clustering scenario
would expect that at $z \sim 2 - 4$ gravity has
not yet had enough time to transfer power on 80 Mpc scales and give
origin to significant over or under-densities.

Numerical simulations in the framework of a CDM model,
including photo-ionization and cooling of the
baryonic component, have reproduced to a remarkable extent
the observed correlation
function and also its dependence on the column density of the
lines \cite{mucket96}.

In this way the observed clustering properties
(in particular the correlation between clustering amplitude and column
density) become part of a new continuity scenario that rests on two
other pieces of evidence: the metallicities of the order $10^{-2}$
observed for Lyman-$\alpha$ clouds with $\log N_{HI} > 14$ 
\cite{cowie95,tytler95} and 
the volume density and cosmological evolution of the same clouds found
to be similar to those of the damped
systems \cite{gial96}. All together this suggests
a physical association between the Lyman-$\alpha$ clouds with $\log N_{HI}
> 14$ and the halos of protogalactic systems.
At low redshift a considerable fraction of the Lyman-$\alpha$ lines
has been observed to be associated with luminous galaxies and the
local large-scale structure 
\cite{lanz95,lebrun96}.

The typical column density below which no clustering of the
Lyman-$\alpha$ lines is observed
($\log (N_{HI}) \sim 13.6$, Fig. \ref{fig1}) 
corresponds to the position of the break in their 
column density distribution 
\cite{gial96,Hu95}, which has been identified with 
the transition from a variety of systems in 
various stages of gravitational infall and collapse
(or even under-densities)
to gas associated with star forming galaxies \cite{anninos96,mucket96}.

Following Heisler, Hogan and White (1989) and Cristiani et al. (1995),
it is possible to derive the three-dimensional spatial autocorrelation
function of the clouds $\xi_r$, from the velocity correlation $\xi_v$,
assuming a Gaussian distribution of the peculiar motions of the clouds
with respect to the Hubble flow and a power-law form for the spatial
correlation function, as observed for galaxies.
$$ \xi_v=\int _0 ^{\infty} H dr~ \xi(r) ~P(v\mid r) \propto
\int_{r_{cl}} ^\infty 
{H dr\over 
\sigma} \left( {r\over r_o}\right) ^{-\gamma} $$
\begin{equation}
\times 
\left \{exp \left [-{(Hr-v)^2 \over
2\sigma ^2}\right]+exp \left [-{(Hr+v)^2 \over 2\sigma ^2}\right]
\right\}
\end{equation}

At small velocity splittings, the correlation scale $r_o$ 
depends mainly on the cloud sizes $r_{cl}$ and on the velocity
dispersion $\sigma$
assumed. 
The very fact that a significant correlation in velocity is observed
results in an upper limit on the cloud proper sizes of the order of $
200 h_{50}^{-1}$
kpc. It is interesting to compare this upper limit with the lower
limits on the cloud dimensions inferred
from the observations of QSO pairs or gravitational lenses
\cite{smette92,smette95,bechto94,dinshaw95}, $r_{cl} \ga 100 h_{50}^{-1}
$ kpc.
Indeed, the trend observed in Fig. \ref{figclustewz} is consistent,
in the framework of a simple hierarchical clustering model predicting
that $r_o$ scales with the redshift according to the relation
$r_o(z) = r_o(0) (1+z)^{-5/3}$,
with large cloud sizes ($r_{cl} \la 200
h_{50}^{-1} $ kpc) and low velocity dispersions
(few tens km/s).

Assuming an index $\gamma =1.77$ for the power-law spatial correlation
function as commonly measured for galaxies, relatively low 
velocity dispersions ($< 100$ km/s)
as suggested by observations of quasar pairs
(e.g. Dinshaw et al. 1994) and cloud sizes of $200 h_{50}^{-1}$ kpc, 
implies, to
reproduce the observed correlation in the velocity space, 
a correlation scale at $z \sim 3$ of $200 - 300  h_{50}^{-1}$ 
comoving kpc.  This amplitude 
extrapolates, according to the $(1+z)^{-5/3}$ scaling,
to $2-3 h_{50}^{-1} $ Mpc at the present epoch, 
to be compared with the values
$10$ and $7.5 h_{50}^{-1}$ Mpc 
observed for present day optical \cite{lappa88} and IRAS \cite{fisher94}
galaxies, respectively.

A spatial correlation 
function with a correlation scale $r_o(z=0) \sim 3$ Mpc,  convolved with 
a velocity dispersion $\sigma \sim 50$ km/s, 
would result in a correlation function
in the velocity space non dissimilar with
respect to what is observed for CIV systems \cite{womb95,son:cow:96}
if a cloud radius $\sim 15 h_{50}^{-1} $ proper kpc is assumed. 
Such a value appears considerably smaller than what is commonly
inferred for galactic 
halos giving origin to metal-rich absorption systems 
($r \sim 75 h_{50}^{-1}$ kpc, Bergeron 1995), but certainly at least
part of the small-scale correlation observed for CIV systems
is due to the motion of clouds of size $ << 75 $ kpc 
within individual galaxy halos.
The observed correlations in the velocity space might then 
arise from the same spatial correlation function both for 
Lyman-$\alpha$ and metal systems, with different degrees of amplitude
suppression due to the different sizes of the regions giving origin
to the Lyman-$\alpha$ and metal absorption, 
a well-known phenomenon occurring when the cloud sizes are non-negligible with
respect to the correlation scale \cite{bajt95}.

\section{Conclusions}

The analysis of a large database ($\sim 1600$ lines) of Lyman-$\alpha$
absorptions observed at $\sim 30000$ resolution in the spectra of 15
quasars has produced the following results:

\begin{enumerate}
  \item clustering on small velocity scales ($\Delta v \la 300~$ km/s) has
been measured at a $5 \sigma$ confidence level for the full line
sample, which has typically $\log (N_{HI}) \ga 13$, with $\xi_v (100 ~{\rm
km/s}) = 0.2 $
  \item the clustering amplitude increases with increasing column
density of the lines: $\xi_v (100 ~{\rm km/s}, \log (N_{HI}) \geq 13.8) = 0.6
$. No significant clustering is observed for lines with $\log (N_{HI})
\la 13.6$. An extrapolation of this trend to column densities typical
of metal systems is consistent with the clustering observed for CIV lines
  \item a trend of decreasing clustering with increasing redshift is observed
  \item a number of over- and under-densities of lines
(including 4 voids) have been identified on scales of a few tens of
Mpc, confirming at a high confidence level the reality of deviations
from a poissonian distribution of lines on such relatively large
scales. The filling factor of such structures is however low: a few
percent of the total line-of-sight path-length.
\end{enumerate}
Although the TPCF approach adopted in the present paper tends to
underestimate the small-scale clustering amplitudes, whose measured
values have to be regarded as lower limits, the following conclusions
can be drawn:
\begin{enumerate}
  \item the observed clustering properties appear qualitatively
consistent with a scenario of gravitationally induced correlations
  \item simple considerations about the two-point correlation function
in real and velocity space allow to deduce an upper limit to the cloud
sizes $r_{cl} \la 200 h_{50}^{-1} $ proper kpc and a correlation scale
$r_o = 200 - 300 h_{50}^{-1} $ comoving kpc at $z \sim 3$.
  \item a continuity scenario between  Lyman-$\alpha$ and metal
systems is suggested,
with a physical 
association between the Lyman-$\alpha$ clouds with $\log N_{HI}
\ga 14$  and the halos of protogalactic systems.
\end{enumerate}

\section*{Acknowledgments}

We thank J. Bergeron, S. Matarrese, L. Moscardini, R. Vio, S. White, 
for  helpful discussions and P. Andreani
for a critical reading of the original version of the
paper.
SC acknowledges the support of the ASI contract 
94-RS-107.

\end{document}